# Do Small Firms Implement Enterprise Systems Differently? The Case of E-Silk Route Ventures

*Research-in-Progress*

**Chandima Wickramatunga**
Swinburne University of Technology,
Melbourne, Australia
cwickramatunga@swin.edu.au

**Darshana Sedera**
Southern Cross University
Queensland, Australia
darshana.sedera@gmail.com

**Sachithra Lokuge**
RMIT University
Melbourne, Australia
ksplokuge@gmail.com

## Abstract

*The cost effectiveness, ease of learning, connectedness and in-depth analytical capabilities provided through cloud computing technologies, have provided small firms the opportunity to implement enterprise systems (ES), which was reserved only for the much resourceful firms. However, it is evident that small firms are still struggling to attain purported benefits of ES and still find it difficult to manage complexities of ES implementations. Further, limited research has been conducted that discusses ES implementation in small firms. This research is an attempt to further the understanding of ES-implementation of small firms and re-evaluate the applicability of fundamental critical success factors to small firms.*

## Introduction

Implementing an enterprise systems (ES) is considered as a costly initiative, as a result, it was exclusively reserved only to the resourceful firms (Lokuge and Sedera 2017; Sedera 2016). However, with the advancement of cloud computing, ES are now becoming an accessible and affordable technology for small firms (Salim et al. 2015; Walther et al. 2018). While implementation of cloud ES is cost effective for small firms, characteristics like complexity of ES, expertise required for implementation and inexperience of the staff have hindered the adoption of ES (Walther et al. 2015; Walther et al. 2013b).

In recent times, there is an increase in adopting cloud solutions by small firms. Prior literature has focused on ES implementation in large and medium sized firms (Sedera 2011); however, limited research has been conducted in ES implementation in small firms. The unique characteristics of small firms, such as, less number of employees, management structure, high top-management involvement, organizational structure, controls, leadership style, decision-making processes, infrastructure, resources and various other organizational practices (Cagliano et al. 2001), have made the implementation of ES in small businesses different to than what is seen in large enterprises. Such characteristics differentiates the implementation process of ES in small firms. Therefore, it is important to study ES implementation in small businesses as it provides a better understanding of the benefits of ES for small firms and provide an avenue to understand critical success factors for ES implementation in small firms. The overarching research question of this study is to investigate how small firms implement complex cloud ES in their firms? This study will focus on crucial success factors (CSF) that enable the ES implementation of a small firm. While there have been many studies on CSF, most of them focus on ES implementations of medium to large firms. As such, this study is an attempt to understand the CSF for ES implementation in a small firm.

To investigate this phenomenon, this research utilizes a qualitative approach and develop a framework that explains the process of implementing ES in a small firm. The findings of this paper will further the understanding of ES implementation of small firms and invigorate a discussion to rethink the critical success factors for ES implementation in small firms. This paper is structured as follows. Next section provides the background of the research followed by methodology section. Then the case study is explained followed by the analysis section. The conclusion section includes contribution to academia, practitioner contribution, limitations, and future research areas.

## Background

ES are corporate information systems (IS), that offer seamless integration of the information throughout the firm (Davenport 1998; Lokuge and Sedera 2018). These software automates core business functions such as supply chain, finance, human resources and manufacturing (Lokuge and Sedera 2014b). Although ES are designed to increase the productivity within an organization, there are many instances where the adoption of such systems has proven to yield negative outcomes (Sedera and Lokuge 2017; Sedera et al. 2016). With huge investment costs and the complicated nature of the implementation, firms are reluctant to adapt such systems. While it has only been large firms who have implemented such systems, small firms have been driven to implement such systems to overcome challenges and gain a competitive edge.

The development of cloud computing has provided a pathway for small businesses to implement ES in their organizations (Lokuge et al. 2016; Walther et al. 2013b). This technology has allowed companies to store, access and utilize information over clouds which has improved the efficiency of their business processes (Salim et al. 2015; Walther et al. 2013a). While cloud computing and other infrastructure has reduced the barriers for ES adoption, it is also important to note that small firms are yet to fully embrace the benefits of ES. While there is prior literature on SME ES implementations, such work seldom can be applied to ES implementation in small firms. This paper attempts to understand the critical success factors (CSF) that contribute towards a successful implementation of an ES in small firms. As per Nah et al. (2003) there are various factors that determine the success of an ES implementation. The next section provides an overview to CSF of ES implementations.

### *CSF of ES Implementation*

While there are a variety of studies that discusses CSF of ES implementations, a very limited literature discusses the CSF of an ES implementation in a small firm. Below are ten commonly discussed CSF of ES implementations.

**Top management support:** This is one of the most important factors that determine the success of ES implementations (Finney and Corbett 2007). It has direct influence on business planning and goal setting, since decisions such as ES implementations are mainly taken by the owner/s (Lokuge 2015).

**Resource availability/ ES Affordability:** This is a very significant factor since small firms have limited resources (Kurnia et al. 2019). The resources refer not only the financial cost and human resources, but also other factors such as the ability to maintain the system and resource availability for future upgrades as well (Huang et al. 2018; Sedera et al. 2003).

**Communication/ Change management strategy:** A proper, effective change management strategy should be developed in order to make the ES adoption transition smooth (Stefanou 2001). Communication among the staff and the managers is key to successful ES implementation (Lokuge et al. 2018).

**Training and education:** Training and education will ensure smooth transition which in turn will increase the organizational performance (Dezdar and Sulaiman 2011). With proper training, user resistance for change can be reduced (Kurnia et al. 2019). Satisfactory training and education, may assist an organization in building up positive attitudes towards a system (Dezdar and Sulaiman 2011).

**Vendor support and selection:** Selecting a suitable vendor plays an important role in ES implementation of SMEs (Zhang et al. 2005). Nowadays, vendors have pre-configured solutions that are specifically designed for SMEs (Venkatraman and Fahd 2016). However, identifying the correct vendor that provides the services suitable for the firm is important (Lokuge and Sedera 2014c).

**ES strategy:** As per Holland et al. (1999), this captures implementation strategy and it suggests that a skeletal system could be implemented initially, and then extra functionality can be added once the system is operating. Execution of the ES can be done by implementing one module at a time along with the existing system or implementing the whole system at once and this is dependent on the firm's propensity to change (Holland et al. 1999).

**Vision and Plan**: Before the ES implementation, a firm should have a clear plan. A clear vision and plan should be developed, that includes the project objectives, and scope that is in line with the IS strategy (Finney and Corbett 2007; Lokuge and Sedera 2014a).

**Business Process Management:** Firms are required to understand the current business structure and the processes associated with the existing IT system and then relate the existing situation to the processes that is within the ES system (Holland et al. 1999).

**Software localization:** Software localization refers to the adjustments that are made to the ES software package to suit the product to the local contexts (Liang and Xue 2004), such as changing of settings (language and user interface) and parameters that reflects local regulations and compliance requirements (Liang and Xue 2004).

**A balanced team:** The project team should be cross functional and should consists of people with diverse skills (Lokuge et al. 2019). Furthermore, this includes the availability of project champions that play a promotional and an influencing role in the project (Finney and Corbett 2007).

# Research Methodology

The objective of this study is to investigate the process of implementing an ES in a small firm and thereby understand the CSF for ES implementation in a small firm. To explore and to get an in-depth understanding of the phenomenon, a qualitative, case study method was applied (Benbasat et al. 1987). The researchers engaged with the case company for a period of seven months investigating the ES implementation project from their inception to completion. The study used snowballing technique to recruit interviewees (Myers and Newman 2007); it sought two types of informants from the organization: (i) the owner and (ii) the managers. The study commenced with interviews with the owner/ director. In addition, the managers were interviewed to obtain a closer perspective of the ES implementation. The data collection was conducted through 10 semi-structured interviews, totaling 15 person-hours. All interviews followed the same case protocol. Each interview took between 1-2 hours and in most cases, follow-up interviews were conducted for clarification. Several new probing interview questions were added based on the interviews which allowed the emergence of new themes. Majority of the interviews were conducted face-to-face and some through virtual platforms (Zoom meetings), in the English language. Further, to obtain an appropriate degree of reliability, internal documentation were used (Dubé and Paré 2003).

## *The Case Organization*

The company that is used for the analysis is a small firm in Asia, referred to as 'E-Silk Route Ventures (Pvt) Ltd.' The company is a supply chain management company operating in the export industry. It was started in 2014 and currently has a workforce of 15 employees. The company's main focused sectors are Organic Agriculture, Food & Beverage, and the Nutraceuticals. The company takes care of Supply Chains from Sourcing, Manufacturing/processing, Packaging, Labelling, comprising - Total Original Brand Manufacturing (OBM) / Private Labelling, Shipping / Logistics, Clearance (for some destinations) and last-mile delivery/fulfilment. The advancement of cloud computing has stimulated the adoption of an ES in E-Silk Route Ventures (Pvt) Ltd (ESRV). The decision to adopt an ES was solely taken by the owner; a characteristic seen in many small firms. The implementation process took a period of 7 months, from the first step of implementation to going live.

The implementation started with the creation of standard operating procedures (SOPs) and mapping out of business processes. Earlier, ESRV did not have a definitive operating procedure, and the implementation of an ES stimulated the development of SOPs and a process map. The initial stage took 2 months and thereafter, they proceeded towards selecting a system. Cost was the main factor, while there were 6 other sub-factors that affected the selection decision. The employees were given an opportunity to suggest ES that would suit ESRV. Two systems were shortlisted, out of which 'Bitrix24' was selected. Bitrix24 is an ES which consists of an inbuilt customer relationship management (CRM) with task automations, a good human resource information system (HRIS) and project management tools. The ES selection process took 8 weeks and thereafter the company proceeded towards system acquisition. The purchase of the system was carried out overnight and only a team of 3 members were present for the acquisition which was headed by the owner. The team utilized social media to get more information about the selected ES and the implementation process. Initial testing was carried out for a week, which included the managerial level employees testing the system and pre-configuring tasks that would help the company in automating processes. A demonstration to the whole team was carried out after the managerial level employees were successful in configuring the system. The subscription of the legacy system was almost near renewal and therefore the Director decided that there was no need to pay for two separate systems. Hence there was a rush towards the last stages of implementation since the existing system was decided not to be renewed. The next stage included the transfer of data and testing of automations. After successfully transferring data, the system went live. ESRV successfully migrated from a simple project management system to an ES within a period of 7 months.

Although the company was successful with the implementation, they were not able to utilize the system to its full potential since they were not able to fully configure the system as per their needs. The owner decided to obtain external support after 8 months but did not go through the process since it was too costly. Therefore, the owner decided to recruit an IT employee with expertise on ES implementation. The new IT personnel managed the rest of the configuration of the ES in the organization.

*Data Analysis*

Following the guidelines of Klein and Myers (1999), an interpretive data analysis approach was applied. As the sensitizing device for our analysis, prior literature on CSF and the vast amount of literature on this topic was applied. After examination of the data, the researchers developed a comprehensive case description detailing the ES implementation (Walsham 1993). The researchers initiated the data analysis process by summarizing and reviewing the interviews throughout the data collection. The initial analysis of data involved identifying factors that determine the success of each of the implementation phase. Further, the researchers paid close attention to the ES lifecycle model (Markus and Tanis 2000), while remaining open to emerging ideas. The categorized data was then further examined through different phases and the emerging themes were then organized to describe CSF of each of the phase. During this stage, the researchers referred to further details such as company documents and referred to the literature as themes emerged. In comparing the emerging themes to existing literature, the researchers adopt existing labels, concepts, or explanations when there were commonalities. Similarly, new labels, concepts, or explanations were developed when there were opportunities to extend the existing literature. The interviews and the project documents were used to identify the sequence of events and utilized to identify the outcomes of each of the phase. The data analysis process was iterative with a recursive feedback obtaining from CSF literature, the data collected and the emerging interpretations. This iterative process continued until consensus was reached. Once we had the core concepts identified, we interpreted them in the context of existing literature, and we demonstrated these empirical patterns into a conceptual framework.

## Findings

When analyzing the ES implementation at ESRV, the researchers created categories and codes from the theoretical viewpoints of CSF and ES implementation, as well as being open to emerging new constructs through the analysis of data. Through the analysis, four phases were derived that drove the whole ES implementation process. Figure 1 depicts the ES project timeline and the phases derived through the analysis. The four phases identified are: Awareness and Strategy formation, Resource exploration, Value creation and System optimization. The first phase includes the awareness of cloud-based ES availability for small firms and forming a proper strategy for the implementation. In this phase generally company do knowledge search and gain an understanding of ES implementation process. The resource exploration phase includes identifying the existing business processes and selection of the best ES compatible with the available resources. The next phase consisted of value creation to ESRV through the implementation of ES which was then followed by optimization of the system that comprised of seeking external support and recruiting of employees with related ES knowledge. We note that the phases of this 'ES lifecycle of small firms' are somewhat different to those that are commonly used in the literature for larger counterparts (Markus and Tanis 2000).

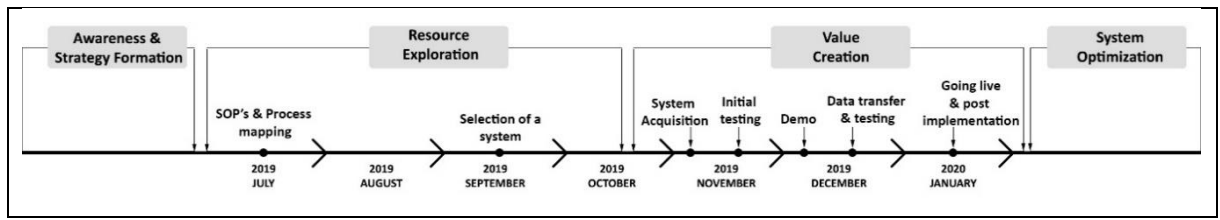

**Figure 1: Timeline of ES Implementation at ESRV**

Each of the CSF identified in the analysis were compared with the implementation of the ES at ESRV. In comparing CSFs with the existing literature, it was found that some CSFs still contributed towards the successful implementation while there were some that were not cited.

Top management support for the whole implementation process had been evident at ESRV. The director had been fully involved and guided the whole implementation process. *"If not for the direction of the owner, we would not be able to implement the ERP successfully"* – The Supply Chain executive.

Resource availability was one of the CSF that was highlighted in the analysis. As with many ES implementations, cost is an important determinant for the software selection. Even though, the expert knowledge on ES implementation is considered important (Sedera and Lokuge 2017), in this case, it did not play an important role.

Communication and effective change management was an important factor throughout the process. Before ES was brought into the company, the director had organized a special meeting to inform the employees about technological developments and how cloud-based systems enabled companies (especially small firms) to implement systems/ software on a subscription basis without having on-premises servers. Even after the first phase was initiated, the whole company was kept informed of the progress. *"The HR manager and I made sure to explain what an ES is to the whole company, and the benefits of the implementation of such a system"* – The owner.

The ESRV staff did not get any training and education for using ES. *"No one in the company had prior experience or training in ES implementation"* – The R&D manager. The owner had decided that they would try to implement the system by themselves by self-learning through YouTube. Furthermore, provision of ES training would take time and be costly. *"We mainly watched the tutorials that were uploaded by the vendor on YouTube. We did not receive any prior training or education. We were asked to learn about Bitrix24 by ourselves"* – The Supply Chain executive.

Vendor support had been a critical factor for ESRV. Since ESRV did not have any prior training or seek for external support, they were dependent on vendor support. The paid version had 24/7 support and ESRV had often sought support during the configuration process. *"We very much relied on having support from the vendor"* – The owner.

Vendor selection was an important factor for ESRV. There were several decisive criteria for selecting a proper vendor such as Cost, Cloud-based, Functionalities, Reviews, Organizational fit, Difficulty and Vendor support. *"The employees were given an opportunity to do a study and come up systems that could be implemented in a small business."* – The Supply Chain executive.

The initial ES strategy was formed during the initial stage. The owner had carried out a feasibility analysis to see whether the ES implementation would assist ESRV to achieve innovation. *"I wanted to see whether an ES was the right way to face future competition and therefore I carried out an analysis."* – Owner.

The owner had a clear vision and strategy as to why they should implement an ES. When ES implementation was under discussion, ESRV started looking at their existing business processes and possible changes that will incur. This was the initial step that was taken in the implementation process. Some of the existing processes were re-structured which made every function streamlined. This helped ESRV to identify what process could be automated. *"The first step of the implementation was to map out all of the business processes. Some processes were kept as they were, but some were changed to make all the functions more streamlined"* – The Supply Chain executive.

The project team consisted of only 3 individuals, i.e., the owner, the research and development manager and the supply chain executive. Although they consisted of basic knowledge on IT, they did not have much knowledge on ES implementation. Hence a balanced team was not evident in this case. *"I did not want to make the implementation complex. I selected 2 other employees for the project. None of them had any experience before, but they were very eager for the implementation"* – The owner.

## Conclusion

The objective of this research-in-progress paper was to understand an ES implementation of a small firm, specifically to derive the applicability of CSF identified through the literature. The topic is of significant value to the contemporary IS research, given that there is a growing number of small firms opting for ES. The study makes several preliminary observations. First, it identifies an ES lifecycle for small firms – where it introduced some specificity to the context of ES. While some phases resemble the traditional ES lifecycle of large firms, phases like 'awareness' provides a unique perspective to the journey of a small firm. For example, in the 'awareness' phase, ESRV looked for inspirations from the 'wider environment,' published cases of large firms and overseas competitors. Considering that there is very little information in the ES literature as to how small firms go about implementing an ES, the primary question of 'are the CSFs derived from large ES implementation studies be any different in small enterprises?' arises. As summarized in Figure 2, while most of the 'factors' are the same, the initial findings show that (i) the sequence in which the CSFs occur, (ii) the specific role that they play in each phase of the lifecycle and the (iii) stakeholder who possess it changes dramatically.

For example, ESRV study demonstrated that the 'implementation plan,' which is typically reserved for the top/middle management of a company, was driven by the software vendor. Similarly, the 'ES strategy' is driven by the employees of ESRV, rather than being led by the top management. The study also identified the impact of resource constraints in small firms and how they impact the implementation process. For example, from Figure 2 it is evident that in ESRV, the emphasis was to 'implement' the system – rather than planning to seek complete value from it. As such, much less emphasis was placed on 'value creation' and 'system optimization.'

We acknowledge the use of single case for the study and the limitations in generalizing the preliminary results. While the above findings are at the initial stage, further studies are underway to confirm these findings and strengthen them. Further cases will be assessed to strengthen the findings and a quantitative study will be conducted to generalize these findings.

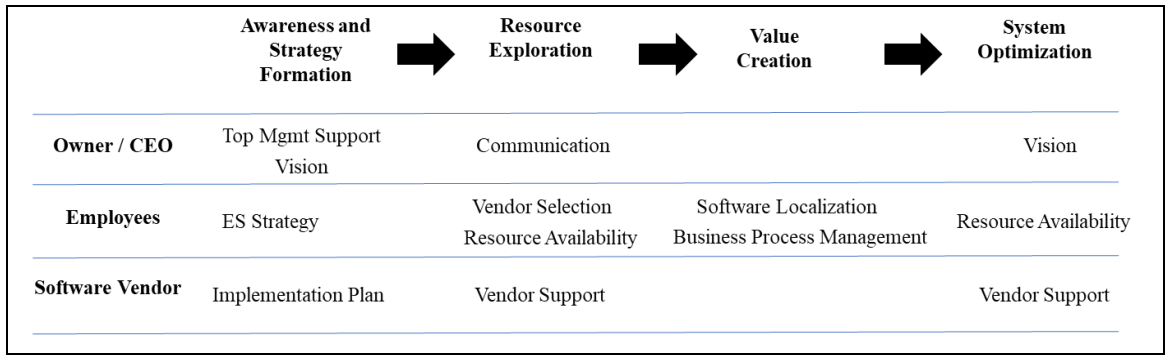

**Figure 1. The Preliminary Framework**

# References

Benbasat, I., Goldstein, D.K., and Mead, M. 1987. "The Case Study Research Strategy in Studies of Information Systems," *MIS Quarterly* (11:3), pp 369-386.

Cagliano, R., Blackmon, K., and Voss, C. 2001. "Small Firms under Microscope: International Differences in Production/Operations Management Practices and Performance," *Integrated Manufacturing Systems* (12:7), 12/01, pp 469-482.

Davenport, T. 1998. "Putting the Enterprise into the Enterprise System," in: *Harvard business review*. pp. 121 - 131.

Dezdar, S., and Sulaiman, A. 2011. "The Influence of Organizational Factors on Successful Erp Implementation," *Management Decision* (49), 06/28, pp 911-926.

Dubé, L., and Paré, G. 2003. "Rigor in Information Systems Positivist Case Research: Current Practices, Trends, and Recommendations," *MIS Quarterly* (27:4), pp 597-636.

Finney, S., and Corbett, M. 2007. "Erp Implementation: A Compilation and Analysis of Critical Success Factors," *Business Process Management Journal* (13), 06/12.

Holland, C., Light, B., and Gibson, N. 1999. *A Critical Success Factors Model for Enterprise Resource Planning Implementation*.

Huang, G., Kurnia, S., and Linden, T. 2018. *A Study of Critical Success Factors for Enterprise Systems Implementation by Smes*.

Klein, H.K., and Myers, M.D. 1999. "A Set of Principles for Conducting and Evaluating Interpretive Field Studies in Information Systems," *MIS Quarterly* (23:1), pp 67-93.

Kurnia, S., Linden, T., and Huang, G. 2019. "A Hermeneutic Analysis of Critical Success Factors for Enterprise Systems Implementation by Smes," *Enterprise Information Systems* (13:9), 2019/10/21, pp 1195-1216.

Liang, H., and Xue, Y. 2004. "Coping with Erp-Related Contextual Issues in Smes: A Vendor's Perspective," *The Journal of Strategic Information Systems* (13:4), 2004/12/01/, pp 399-415.

Lokuge, K.S.P. 2015. "Agile Innovation: Innovating with Enterprise Systems," in: *Information Systems School*. QUT ePrints: Queensland University of Technology.

Lokuge, S., and Sedera, D. 2014a. "Deriving Information Systems Innovation Execution Mechanisms," *Australasian Conference on Information Systems*, Auckland, New Zealand: AIS.

Lokuge, S., and Sedera, D. 2014b. "Enterprise Systems Lifecycle-Wide Innovation," *Americas Conference on Information Systems (AMCIS 2014)*, Savannah, Georgia: AIS.

Lokuge, S., and Sedera, D. 2014c. "Enterprise Systems Lifecycle-Wide Innovation Readiness," *Pacific Asia Conference on Information Systems*, Chengdu, China: AIS.

Lokuge, S., and Sedera, D. 2017. "Turning Dust to Gold: How to Increase Inimitability of Enterprise System," *Pacific Asia Conference on Information Systems*, Langkawi, Malaysia: AIS.

Lokuge, S., and Sedera, D. 2018. "The Role of Enterprise Systems in Fostering Innovation in Contemporary Firms," *Journal of Information Technology Theory and Application (JITTA)* (19:2), pp 7-30.

Lokuge, S., Sedera, D., and Grover, V. 2016. "Thinking inside the Box: Five Organizational Strategies Enabled through Information Systems," *Pacific Asia Conference on Information Systems*, Chiyai, Taiwan: AIS.

Lokuge, S., Sedera, D., Grover, V., and Xu, D. 2019. "Organizational Readiness for Digital Innovation: Development and Empirical Calibration of a Construct," *Information & Management* (56:3), pp 445-461.

Lokuge, S., Sedera, D., and Perera, M. 2018. "The Clash of the Leaders: The Intermix of Leadership Styles for Resource Bundling," *Pacific Asia Conference on Information Systems*, Yokohama, Japan: AIS.

Markus, M.L., and Tanis, C. 2000. "The Enterprise Systems Experience-from Adoption to Success," *Framing the Domains of It Management* (173:2000), pp 207-173.

Myers, M.D., and Newman, M. 2007. "The Qualitative Interview in Is Research: Examining the Craft," *Information & Organization* (17:1), pp 2-26.

Nah, F.F., Zuckweiler, K.M., and Lau, J.L. 2003. "Erp Implementation: Chief Information Officers' Perceptions of Critical Success Factors," *International Journal of Human-Computer Interaction* (16:1), pp 5-22.

Salim, S.A., Sedera, D., Sawang, S., Alarifi, A.H.E., and Atapattu, M. 2015. "Moving from Evaluation to Trial: How Do Smes Start Adopting Cloud Erp?," *Australasian Journal of Information Systems* (19), pp S219-S254.

Sedera, D. 2011. "Size Matters! Enterprise System Success in Medium and Large Organizations," in: *Enterprise Information Systems: Concepts, Methodologies, Tools and Applications,* I.R.M. Association (ed.). Pennsylvania, United States: IGI Global, pp. 958-971.

Sedera, D. 2016. "Does Size Matter? The Implications of Firm Size on Enterprise Systems Success," *Australasian Journal of Information Systems* (20), pp 1-25.

Sedera, D., Gable, G., and Chan, T. 2003. "Survey Design: Insights from a Public Sector-Erp Impact Study," *Pacific Asia Conference on Information Systems*, Adelaide, Australia: AIS, pp. 595-610.

Sedera, D., and Lokuge, S. 2017. "The Role of Enterprise Systems in Innovation in the Contemporary Organization," in: *The Routledge Companion to Management Information Systems,* R.G. Galliers and M.-K. Stein (eds.). Abingdon, United Kingdom: The Routledge p. 608.

Sedera, D., Lokuge, S., Grover, V., Sarker, S., and Sarker, S. 2016. "Innovating with Enterprise Systems and Digital Platforms: A Contingent Resource-Based Theory View," *Information & Management* (53:3), pp 366–379.

Stefanou, C.J. 2001. "Organizational Key Success Factors for Implementing Scm/Erp Systems to Support Decision Making," *Journal of Decision Systems* (10:1), 2001/01/01, pp 49-64.

Venkatraman, S., and Fahd, K. 2016. "Challenges and Success Factors of Erp Systems in Australian Smes," *Systems* (4), 05/05, p 20.

Walsham, G. 1993. *Interpreting Information Systems in Organizations*. Chichester: Wiley & Sons.

Walther, S., Sarker, S., Sedera, D., and Eymann, T. 2013a. "Exploring Subscription Renewal Intention of Operational Cloud Enterprise Systems-a Socio-Technical Approach," *European Conference on Information Systems*, Utrecht, The Netherlands: AIS, p. 25.

Walther, S., Sarker, S., Urbach, N., Sedera, D., Eymann, T., and Otto, B. 2015. "Exploring Organizational Level Continuance of Cloud-Based Enterprise Systems," in: *European Conference on Information Systems*. Münster, Germany: AIS.

Walther, S., Sedera, D., Sarker, S., and Eymann, T. 2013b. "Evaluating Operational Cloud Enterprise System Success: An Organizational Perspective," *European Conference on Information Systems (ECIS 2013)*, Utrecht, p. 16.

Walther, S., Sedera, D., Urbach, N., Eymann, T., Otto, B., and Sarker, S. 2018. "Should We Stay, or Should We Go? Analyzing Continuance of Cloud Enterprise Systems," *Journal of Information Technology Theory and Application (JITTA)* (19:2), pp 57-88.

Zhang, Z., Lee, M.K.O., Huang, P., Zhang, L., and Huang, X. 2005. "A Framework of Erp Systems Implementation Success in China: An Empirical Study," *International Journal of Production Economics* (98:1), 2005/10/18/, pp 56-80.